\documentclass[fleqn,10pt]{wlscirep}
\usepackage[utf8]{inputenc}
\usepackage[T1]{fontenc}
\usepackage{lineno}
\usepackage{tabularx}
\usepackage{booktabs}
\usepackage{xurl}
\usepackage{amsmath}

\usepackage{booktabs}
\usepackage{multirow}
\usepackage{graphicx}
\usepackage{array}
\usepackage{xcolor}

\title{GLObal Building heights for Urban Studies (UT-GLOBUS) for city- and street- scale urban simulations: Development and first applications}
\author[1]{Harsh G. Kamath}
\author[1]{Manmeet Singh}
\author[2]{Neetiraj Malviya}
\author[3]{Alberto Martilli}
\author[4]{Liu He}
\author[4]{Daniel Aliaga}
\author[5]{Cenlin He}
\author[5]{Fei Chen}
\author[6]{Lori A. Magruder}
\author[1]{Zong-Liang Yang}
\author[1,7,*]{Dev Niyogi}
\affil[1]{Department of Earth and Planetary Sciences, Jackson School of Geosciences, University of Texas at Austin, Austin, Texas, USA}
\affil[2]{Indian Institute of Tropical Meteorology, Ministry of Earth Sciences, Pune, India}
\affil[3]{CIEMAT, Madrid, Spain}
\affil[4]{Department of Computer Science, Purdue University, West Lafayette, Indiana, USA}
\affil[5]{National Center for Atmospheric Research, Boulder, Colorado, USA}
\affil[6]{Department of Aerospace Engineering and Engineering Mechanics, University of Texas at Austin, Texas, USA}
\affil[7]{Fariborz Maseeh Department of Civil, Architectural and Environmental Engineering, University of Texas at Austin, Austin, Texas, USA}

\affil[*]{corresponding author(s): Dev Niyogi (happy1@utexas.edu)}

\begin{abstract}

We introduce GLObal Building heights for Urban Studies (UT-GLOBUS), a dataset providing building heights and urban canopy parameters (UCPs) for major cities worldwide. UT-GLOBUS combines open-source spaceborne altimetry (ICESat-2 and GEDI) and coarse-resolution urban canopy elevation data with a machine-learning model to estimate building-level information. Validation using LiDAR data from six U.S. cities showed UT-GLOBUS-derived building heights had a root mean squared error (RMSE) of 9.1 meters. Validation of Mean building heights within 1-km² grid cells, including data from Hamburg and Sydney, resulted in an RMSE of 7.8 meters. Testing the UCPs in the urban Weather Research and Forecasting (WRF-Urban) model resulted in a significant improvement (~55\% in RMSE) in intra-urban air temperature representation compared to the existing table-based local climate zone approach in Houston, TX. Additionally, we demonstrated the dataset's utility for simulating heat mitigation strategies and building energy consumption using WRF-Urban, with test cases in Chicago, IL, and Austin, TX. Street-scale mean radiant temperature simulations using the SOlar and LongWave Environmental Irradiance Geometry (SOLWEIG) model, incorporating UT-GLOBUS and LiDAR-derived building heights, confirmed the dataset’s effectiveness in modeling human thermal comfort in Baltimore, MD (daytime RMSE = 2.85°C).  Thus, UT-GLOBUS can be used for modeling urban hazards with significant socioeconomic and biometeorological risks, enabling finer scale urban climate simulations and overcoming previous limitations due to the lack of building information.

\end{abstract}
\begin{document}

\flushbottom
\maketitle

\thispagestyle{empty}


\section*{Background \& Summary}
As urban areas host the majority of the world's population, there is an interest in representing cities in weather and climate models for environmental studies. Heat and air quality are particularly critical urban stressors that many cities globally prioritize. Accurate local simulations of these stressors require detailed 3-D information about the urban environment. For example, Kamath et al., (2023) demonstrated that shade from buildings can significantly affect human thermal comfort\cite{kamath2023human}, while Lewis et al., (2024) highlighted the impact of 3-D urban structures on smoke dispersion\cite{lewis2024fire}. Typically, light detection and ranging (LiDAR) surveys and satellite photogrammetric digital surface models (DSM) are employed to obtain building information for urban simulations. However, LiDAR surveys are limited in global availability and spatial coverage, while high-resolution photogrammetric DSM data often suffer from noise and require additional processing to produce accurate DSMs\cite{Wang2021}. Additionally, these datasets are generally not open-source, except for LiDAR data covering the contiguous United States and a few other specific locations worldwide. Moreover, these datasets are not directly compatible with urban models. Therefore, a framework is needed to generate building information for urban studies. These studies may involve investigating biometeorology, urban energy consumption, air quality, and the impacts of large-scale events such as thunderstorms on cities.

Present-day models effectively capture the variables necessary to study the urban climate at city- and street-scales. However, a significant challenge lies in providing the building dataset required as input for the models. These models include urban canopy models (UCMs) such as those implemented in the Weather Research and Forecasting (WRF-Urban) model\cite{Chen2011}, urban energy balance models such as the Surface Urban Energy and Water Balance Scheme (SUEWS)\cite{sun2019python}, and outdoor thermal comfort models like the SOlar and LongWave Environmental Irradiance Geometry (SOLWEIG)\cite{Lindberg2008}. For cities lacking detailed building data, WRF-Urban and SUEWS often use the concept of local climate zones (LCZ)\cite{Stewart2012}. This approach categorizes built-up areas into urban zones based on urban canopy parameters (UCPs) such as average building height, plan area fraction, and sky-view factor, which are relatively uniform within an urban block. The models then use a predefined look-up table of UCPs to represent the morphology of each urban zone. However, this method can only broadly represent a city since UCPs for LCZs vary across different cities, and the UCPs table does not accurately reflect the specific characteristics of an urban block or the entire city\cite{Patel2023}. 

An early attempt to incorporate detailed UCPs into WRF-Urban was made through the National Urban Database and Access Portal Tool (NUDAPT) dataset\cite{Ching2009}. This dataset includes LiDAR-derived UCPs for 44 cities in the US. A recent effort to expand this concept globally has led to the development of the World Urban Database and Access Portal Tool (WUDAPT)\cite{Ching2018}. WUDAPT currently provides level-0 or LCZ information globally but lacks detailed UCPs specific to individual cities\cite{Ching2018}. 

City-scale simulations using WRF-Urban have previously highlighted the impacts of incorporating detailed UCPs on modeling air temperature, energy consumption, and heat mitigation strategies. Salamanca et al., (2011)\cite{Salamanca2011} compared observed near-surface air temperatures (T2M) with simulations using bulk, single-layer\cite{Kusaka2001}, and multi-layer\cite{Martilli2002,Salamanca2010} UCMs, utilizing both table-based and NUDAPT-derived UCPs. Their results indicated that detailed urban morphology significantly improved model accuracy. Additionally, the multi-layer UCM demonstrated greater sensitivity to UCPs, primarily due to anthropogenic heat emissions from air conditioning. Papaccogli et al., (2021)\cite{Papaccogli2021} emphasized the influence of building height and plan area fraction on winter heating demand, showing that incorporating detailed model inputs captured the spatial pattern of building energy consumption in simulations using the multi-layer model. Detailed UCPs are also crucial in simulating urban heat mitigation strategies, such as green and cool roofs, as highlighted by Zonato et al., (2021)\cite{Zonato2021}. However, due to constraints in data availability, heat mitigation strategies are often simulated using default LCZ-based UCPs, as noted by Tan et al., (2023)\cite{Tan2023}.

Street-scale urban climate modeling is significantly influenced by the 3-D urban environment. For instance, shading due to the buildings and vegetation canopies affects human thermal stress. Thus, building-level information is essential for accurately modeling heat health assessment metrics such as mean radiant temperature ($T_{MRT}$). For street-scale modeling, building information is primarily sourced from LiDAR surveys or OpenStreetMap (OSM). However, the OSM dataset may have incomplete building heights and footprints\cite{Bernard2022}.

To address the data availability challenges in city- to street-scale urban modeling, we resort to machine learning (ML) approaches. Over the past decade, ML has achieved remarkable success across various domains, particularly in computer vision and tabular data prediction\cite{Reichstein2019}. Predicting building heights and 2-D footprints has become a relevant problem within ML, and decision tree-based models like random forests (RF)\cite{Breiman2001} have been used on tabular datasets to predict missing building heights using data from OpenStreetMap\cite{Bernard2022}.

This data description paper introduces a new dataset named GLObal Building Heights for Urban Studies (UT-GLOBUS), which provides 'level-of-detail-1' building information and UCPs for major cities across all habitable continents. The building-level information available in vector file format has individual building polygons with height attribute in meters above ground level. The vector files are suitable for visualization using GIS platforms, while the UCPs are provided in a binary file format compatible with the WRF preprocessing system. The primary objective of UT-GLOBUS is not to precisely predict the height and footprint of individual buildings, but to offer a dataset for modeling applications from city- to street-scales. Specifically, it is designed to meet the requirements for deriving UCPs for the multi-layer UCM in the WRF-Urban model and for providing building heights for the SOLWEIG model. Consequently, the UT-GLOBUS dataset is mainly targeted at users with prior experience with the WRF-Urban and SOLWEIG models, although it should be of value for other engineering, weather, and urban planning applications.

\section*{Methods}
This section describes the datasets utilized and methods used for data processing, generation, and validation and provides an overview of the application areas of the UT-GLOBUS dataset.

\subsection*{Data acquisition and processing}
Datasets for model training were acquired for urban blocks that represent high-rise, mid-rise, and low-rise buildings and their combinations. By capturing the diverse range of building sizes, the model can generalize and make better predictions for building heights.

\textbf{LiDAR:} Building heights derived from the United States Geological Survey (USGS) 3DEP point cloud data served as the ground truth. To facilitate semantic segmentation, feature class point labeling was conducted, followed by the rasterization of the building height layer at a 1-meter spatial resolution. The resolution of the raster was determined based on the density of LiDAR point clouds per square meter. In addition to building DSM, a digital elevation model (DEM), representing the bare earth, was generated from the LiDAR point cloud data. By subtracting the building DSM from the DEM, normalized DSM (nDSM) was obtained. 

\textbf{Spaceborne elevation datasets:} To acquire the spaceborne nDSM at a 30-meter spatial resolution, the JAXA ALOS/PRISM near-global stereo DSM\cite{Tadono2015} was subtracted from the DEM from Shuttle Radar Topography Mission. It's important to note that the ALOS mission concluded in 2011, potentially missing recent urban development data. To mitigate this limitation, data from the spaceborne photon counting altimeters Ice, Cloud and Land Elevation Satellite-2 (ICESat-2) and Global Ecosystem Dynamics Investigation (GEDI) were incorporated. ICESat-2 and GEDI provide elevation data along their orbital tracks, which initially offered sparse spatial coverage. To enhance coverage, data from multiple orbital passes of each sensor were integrated. Further details on the data sparsity issue and the impact of integrating ICESat-2 and GEDI on building height estimation are illustrated in Figures S.1.1 and S.2.1, respectively, in the supplementary material.

ICESat-2 ATL03 elevation product\cite{Neuenschwander2019} provides elevation measurements at a resolution of 0.7 meters along the track for each of its six lasers. However, since ATL03 provides raw photon data, the ICESat-2 ATL08 product was utilized to classify these photons into ground and top-of-canopy categories. ATL08 is a higher-level product that offers ground/canopy classification and elevation at a length scale of 100 meters. Only ATL03 photons classified with medium and high confidence levels were retained for this study. Within a 30-meter regular grid, the maximum canopy top height and ground elevation were subtracted to derive the ICESat-2 above ground level height (AGLH). On the other hand, GEDI has an along-track resolution of 60 meters and an across-track resolution of 600 meters with 8 lasers. For this study, the GEDI top of canopy RH98 product was used. The ICESat-2 and GEDI above ground level height layers were converted into point vector layers and then merged. This combined layer underwent triangular interpolation to produce a raster at a spatial resolution of 30 meters. Subsequently, this interpolated raster was integrated with the ALOS nDSM. This integration replaced ALOS nDSM pixels lacking height values with corresponding interpolated raster pixels derived from ICESat-2 and GEDI data. This process resulted in the creation of the spaceborne normalized nDSM.

Even with the utilization of ICESat-2 and GEDI products, the data sparsity issue limited the comprehensive coverage of recent urban sprawl for many cities. To further address this challenge, the World Settlement Footprint (WSF) 3-D dataset\cite{esch2022world} was employed. This dataset provides average building heights globally at a coarser spatial resolution of ~90 meters. The height of the spaceborne nDSM was adjusted using the WSF 3-D dataset using a scaling factor to further capture the recent urban sprawl and high-rise buildings.
\begin{equation}
\mathit{Scaling \; factor} \; = \; \frac{\mathit{WSF \; 3\text{-}D}_{90 \ \mathit{meters}}}{\mathit{Spaceborne \; nDSM}_{90 \ \mathit{meters}}}
\end{equation}

\begin{equation}
\mathit{Adjusted \; spaceborne \; nDSM}_{30 \ \mathit{meters}} \; = \; \mathit{Spaceborne \; nDSM}_{30 \ \mathit{meters}} \; \times \; \mathit{scaling \; factor}
\end{equation} where $spaceborne\,nDSM_{90\,meters}$ is the nDSM obtained by spatially aggregating the 30-meter resolution spaceborne nDSM to 90 meters.

\textbf{Population:} For predicting building heights, the population density was incorporated due to its correlation with mean building heights, as demonstrated by Frantz et al., (2021)\cite{Frantz2021}. Higher population density within a smaller building footprint suggests taller buildings. For this study, we utilized the globally available open-source Landscan population density data (people per grid cell) for the year 2020, which has a spatial resolution of approximately 1 km$^2$ \cite{dobson2000landscan}. In our analysis, the linear relationship between population density, building footprint area, and mean building height was found to be statistically significant with a p-value <<0.05. 

The population density data, initially available at a 1 km spatial resolution, underwent smoothing using cubic convolution methods to reduce noise. Subsequently, it was reprojected onto the 30-meter resolution grid of the spaceborne nDSM. When predicting building heights for cities outside the United States, a simple correction factor was applied to adjust for differences in population density as our ML model was trained over the cities in the US. This correction factor was determined by calculating the ratio of the population of a reference US city to the population of the city being processed.

\textbf{2-D building footprints:} For 2-D building footprints, UT-GLOBUS utilizes data sourced from OSM (\url{http://api.openstreetmap.org/}), Google (\url{https://sites.research.google/open-buildings/}), and Microsoft (\url{https://www.microsoft.com/en-us/maps/bing-maps/building-footprints}). These datasets are combined to achieve nearly complete coverage: Microsoft provides footprints for the Northern Hemisphere, while Google covers the Southern Hemisphere. In cases where cities lack these datasets or have a significant number of missing footprints across all three sources, UT-GLOBUS employs two generative deep learning-based methods for building feature extraction: Generative Building Feature Estimation (GBFE)\cite{he2023generative} and Urban Layout Generator (GlobalMapper)\cite{he2023globalmapper}.

GBFE utilizes a two-stage approach. Initially, it performs semantic segmentation of satellite imagery using a U-NET architecture to identify building regions. In the second stage, a generator model is trained to generate precise building footprints based on the segmented regions. GBFE excels in extracting footprints from coarse resolution (e.g., 3 meters) or blurry satellite imagery that is widely available globally. Its performance surpasses existing methods, as evidenced by comprehensive documentation\cite{he2023generative}. Figure 1a below demonstrates GBFE's performance against ground truth in an urban block in Vienna, Austria. GlobalMapper generates building footprints given road networks and a fraction of existing building footprints (from OSM, GBFE, etc.). This method addresses data equity issues in medium and small cities with limited data accessibility. These approaches enable UT-GLOBUS to enhance data coverage, particularly in regions where traditional dataset availability is limited or incomplete.

If OSM footprints are used, additional processing steps involve filling in any gaps within the building polygons, and in situations where multiple adjacent buildings share a common wall, these buildings are merged into a single polygon. This merging process helps simplify the representation of the buildings and ensures that the resulting polygons are more suitable for calculating UCPs. Figure 1b below visually illustrates the differences between the original OSM data and the processed data after applying simplifications to the building footprints. It is important to note that the existing building heights from the OSM dataset are retained in UT-GLOBUS whenever they are available. The Microsoft and Google building footprint data do not require any additional processing.

\subsubsection*{Training}
We used a random forest model to predict building heights, and the training and validation process is shown in Figure 2. Several attributes were assigned to each building's 2-D footprint polygon for training and validation. These attributes are listed below:

\textbf{Ground truth for training:} Using the LiDAR-based nDSM building height (in meters), the height attribute for each building 2-D footprint polygon was assigned and used as the target variable for training.

\textbf{Spaceborne above ground level height:} The above ground level height is assigned to each building 2-D footprint polygon using the WSF 3-D adjusted spaceborne nDSM, providing primary information about building heights.

\textbf{Population density:} The average population density at each building 2-D footprint polygon was assigned as an attribute to capture the number of people living within the vicinity of the building.

\textbf{Area:} The area of each building polygon is calculated and assigned as an attribute. This attribute represents the spatial extent of the buildings.

The training dataset utilized in this study encompasses buildings from six cities in the USA: New York City, Philadelphia, Boston, Los Angeles, and San Francisco, totaling approximately 268,000 buildings. To train the random forest regression model, 80\% of the data points were randomly selected and used for model fitting, while the remaining 20\% were reserved for validation. A hyperparameter tuning process was conducted to optimize the RF model's performance. This involved exploring various combinations of parameters and evaluating the model's effectiveness using 3-fold cross-validation. The RF model employed 240 estimators, required at least 12 samples to split a node, and set a minimum of 2 samples to form a leaf. The number of features considered at each split was determined as the square root of the total number of features. Each tree in the RF ensemble had a maximum depth of 50, and bootstrap sampling was enabled. For more detailed information on the hyperparameters used, please refer to the documentation available at \url{https://scikit-learn.org/stable/modules/generated/sklearn.ensemble.RandomForestRegressor.html}

\subsubsection*{Method for technical validation}
In evaluating UT-GLOBUS, comparisons with LiDAR-derived building heights were conducted using several metrics: root mean squared error (RMSE), mean bias error (MBE), and coefficient of determination (\(R^2\)). Beyond building heights, UT-GLOBUS provides UCPs essential for the multi-layer urban model in WRF-Urban at a resolution of 1 \(km^2\). The UCPs include the plan area fraction (\(\lambda_p\)), building surface to plan area ratio (\(\lambda_b\)), and area-averaged building height (\(h_a\)) along with the building height histogram with 5-meter bin size. Since nearest neighbor interpolation is used to assign the UCPs to urban grids in the WRF preprocessing system at the native resolution of the model, we have calculated UCPs using a 300-meter sliding kernel with 1 \(km^2\) area to ensure that the maximum geographical mismatch during interpolation is limited to <= 300 meters. This UCP calculation process is visually shown in Figure S.3.1 in the supplementary material.  \(\lambda_p\) is defined as the ratio of the total building footprint area \(A_f\) to the total grid area \(A_t\) considered. 
\begin{equation}
    \lambda_p = \frac{A_f}{A_t}
\end{equation}

The building surface to plan area ratio \((\lambda_b)\) is defined as the ratio of the total building surface in the urban canopy layer (area of roofs and walls) to the grid area considered.
\begin{equation}
    \lambda_b = \left(\frac{A_r + A_w}{A_t}\right) = \left(\frac{A_r + (P \times H)}{A_t}\right)
\end{equation}

where \(A_r ( = A_f )\) and \(A_w\) are the roof and wall areas, respectively. Since UT-GLOBUS generates LoD-1 buildings, \(\lambda_b\) can be simplified using the building footprint perimeter \(P\) and building height \(H\). The area-averaged building height \((h_a)\) is defined as
\begin{equation}
    h_a = \frac{\left(\sum_{i=1}^N A_i h_i\right)}{\left(\sum_{i=1}^N A_i\right)}
\end{equation} where \(h_i\) and \(A_i\) are the height and area of \(i\)th building within a grid with \(N\) buildings. In addition to the metrics used for validating building heights, we also calculated the spatial correlation coefficient (SC) to assess the accuracy of area-averaged building heights.

The selection of cities for validating the UT-GLOBUS dataset was based on the availability of open data sources. To evaluate the performance of building height prediction on data points outside the training and validation sets, a separate testing dataset was compiled. This testing dataset included 123,020 buildings from six cities: Atlanta, Austin, Chicago, Houston, Pittsburgh, and San Antonio. For the validation of area-averaged building heights, we utilized the same six cities in the USA, along with two additional cities: Hamburg, Germany, and Sydney, Australia. These cities were included to ensure robust evaluation across different urban contexts and geographic regions.

\subsubsection*{Areas of data usage}
UT-GLOBUS provides vector files containing building 2-D footprint polygons along with their heights above ground in meters as an attribute. This facilitates the computation of necessary UCPs for the multi-layer UCM for city-scale weather simulations. The UCPs were calculated and transformed into a binary file format suitable for ingestion into the WRF preprocessing system. Table 1 summarizes the UCPs derived from UT-GLOBUS that apply to the UCMs in WRF-Urban. A detailed procedure for integrating UT-GLOBUS data into WRF-Urban is available in section 4 of the supplementary material. It is important to note that although additional UCPs such as sky view factor and roughness lengths can be derived from UT-GLOBUS, we only provide UCPs required for the multi-layer UCM.

The UT-GLOBUS dataset has applications for modeling urban climate at city-to-street scales. To evaluate its added value at the city scale, we compared WRF-Urban simulations of T2M and land surface temperature (LST) against observations and satellite-derived products using UT-GLOBUS and default LCZ UCPs. Additionally, we compared simulated T2M under heat mitigation strategies and urban energy consumption using UT-GLOBUS and default LCZ-based UCPs. Although the heat mitigation and urban energy consumption experiments lack experimental data, they offer insights into the performance and potential benefits of UT-GLOBUS. For street-scale human thermal comfort simulations, we compared $T_{MRT}$ simulations using UT-GLOBUS with a control case using LiDAR-derived building heights.

\section*{Data Records}
The dataset containing building heights in vector file format and urban canopy parameters (UCPs) in WRF pre-processing system compatible binary file format can be accessed on Zenodo\cite{kamath_2024}. We are also supplying a city coverage vector file showing the cities where UT-GLOBUS data is available. The cities coverage file \texttt{coverage\_xxxx.gpkg} can be opened in platforms like ArcGIS and QGIS to view the complete list of cities along with the geographic extents. Here, ‘xxxx’ refers to the name of the geographic region, e.g., Asia or the USA. The UT-GLOBUS building vector files employ the Universal Transverse Mercator (UTM) projection. These vector files can be rasterized and then used in SUEWS and SOLWEIG models. The vector files are compatible with QGIS and ArcGIS and can be imported for analysis using programming languages such as Python. Additionally, we provide the urban fractions calculated using the ESA world cover dataset (\url{https://esa-worldcover.org/en}) for the WRF-Urban model in binary file format. 

\section*{Technical Validation}
\subsubsection*{Validation of building heights}
A scatter plot presented in Figure 3 provides an overview of the random forest model performance for predicting the building heights across different cities and urban blocks for both the validation and testing datasets. The RMSE values for the validation and testing datasets were 5.4 and 9.1 meters, respectively. These RMSE values are consistent with expectations, considering the ALOS DSM, a predictor in the RF model, has an RMSE of about 4 meters\cite{Tadono2015}. The MBE for validation and testing samples were -0.06 meters and 0.1 meters, respectively. A box plot in Figure 3 illustrates the bias in predictions by grouping heights into 10-meter bins. The analysis reveals a positive bias that increases with building height, indicating the random forest model tends to overestimate building heights and shows reduced performance for predicting heights of taller buildings. This aligns with the positive MBE observed for the testing dataset. The overestimation may stem from the WSF 3-D dataset providing area average heights that are larger than actual, and the RF model's use of lower-resolution input datasets may contribute to reduced accuracy in predicting taller building heights. The overestimation of building heights can result in an overestimation of air temperature and urban energy consumption in u-WRF simulations, owing to anthropogenic heat release and building cooling demand. For street-scale simulations, this also likely leads to an overestimation of shaded areas. Despite these challenges, the model demonstrated good performance in predicting the interquartile range of building heights, indicating overall effectiveness.

\subsubsection*{Validation of UT-GLOBUS-derived urban canopy parameters (UCPs)}
Figure 4 below compares area-averaged building heights from UT-GLOBUS and Li et al., (2020)\cite{li2020developing} with LiDAR-based ground truth data at a 1 km$^2$ resolution for Austin, Texas. It also includes scatter plots of UT-GLOBUS and Li et al. (2020) heights against LiDAR-based heights. Table 2 presents the RMSE, MBE, \(R^2\), and spatial correlation coefficient for all cities where validation was performed. Spatial validation for the cities listed in Table 2 is detailed in Figures S.5.1 to S.5.7 in the supplementary material. The validation statistics presented in Table 2 indicate that UT-GLOBUS outperforms the existing dataset across the USA, while the dataset by Frantz et al., (2021)\cite{Frantz2021} excels in Hamburg, Germany when compared to the European Space Agency (ESA) dataset (\url{https://land.copernicus.eu/en/products/urban-atlas/building-height-2012}). The comparison of UT-GLOBUS heights with the dataset by Lipson et al., (2022)\cite{lipson2022transformation} for Sydney, Australia showed performance that is similar to cities in the US.

Additionally, Figure 5 shows the spatial comparison of UCPs calculated from equations 3-5 using the UT-GLOBUS and Spanish building data inventory datasets for Madrid, Spain. The RMSE for the area-averaged building heights was 5.53 meters, indicating that UT-GLOBUS performance for Madrid is comparable to other cities in Table 3. The high spatial correlation score of 0.8 for \(\lambda_b\) confirms that UT-GLOBUS performed well in predicting building footprints, in addition to height predictions, as \(\lambda_b\) considers both height and footprint area as seen from equation 4.

\section*{Usage notes }
\subsubsection*{City-scale WRF-Urban simulations}
We utilized WRF-Urban model version 4.4\cite{Chen2011}. The detailed model set-up is provided in section 6 of the supplementary material. The city-scale experiments we conducted are outlined in Table 4. 

\textbf{Land Surface Temperature (LST) simulation using UT-GLOBUS:} The WRF-Urban simulated LST was compared with the ECOSTRESS ECO2LSTE product\cite{Hulley2022}. ECOSTRESS, mounted on the International Space Station, provides LST data at ~70 meters spatial resolution and can capture LST at different times of the day over a specific location with a return cycle of ~3-5 days due to its non-sun-synchronous orbit. For comparison with WRF-Urban simulations, a diagnostic variable was added in multi-layer UCM to isolate LST contributions from roofs and streets. However, spaceborne sensors typically do not provide LST for the entire urban surface or exclusively for roofs and streets\cite{Voogt1997} (horizontal surfaces). Urban LST data from satellite sensors are influenced by urban morphology (shading effects and reduced sky-view factor) and large off-nadir sensor view angles\cite{anderson2021interoperability} (affecting the spatial resolution of the retrieved LST).

Figure 6 shows ECOSTRESS and simulated LST data for Chicago at ~12 PM. For the comparison in Figure 6, ECOSTRESS data were not filtered for lower or near-nadir view angles to obtain LST of only horizontal surfaces due to the availability of clear-sky scenes. Therefore, it is crucial to recognize that ECOSTRESS LST data in Figure 6 cannot be directly compared to WRF urban simulations representing the LST of horizontal surfaces exclusively. However, given the lack of alternative sensors providing LST of horizontal surfaces at various times of the day, ECOSTRESS was utilized. At mid-day, the LST of the entire urban surface is lower compared to the LST of horizontal surfaces (measured at nadir), while at night, it is higher\cite{oke2017urban}. This is because roof surfaces are cooler than the complete urban surface at night and warmer during the day. A similar comparison for Houston at ~12 AM is shown in Figure S.7.1 in the supplementary material, with validation statistics for both cities presented in Table 5.

The spatial comparison presented in Figure 6 and statistics shown in Table 5 demonstrates that UT-GLOBUS is capable of capturing LST more accurately than the LCZ approach during daytime in Chicago ($RMSE_{UT-GLOBUS}$ = 2.27 K and $RMSE_{LCZ}$ = 3.03 K). However, in Houston, the RMSE using the LCZ approach is lower during nighttime ($RMSE_{UT-GLOBUS}$ = 1.22 K and $RMSE_{LCZ}$ = 1.1 K). However, $MBE_{UT-GLOBUS}$ was 0.51 K compared to -0.66 K for $MBE_{LCZ}$. 

\textbf{Air temperature (T2M) Simulation using UT-GLOBUS:} A heat mapping campaign was conducted in Houston on August 7th, 2020, with 84 volunteers collecting data along 32 routes. To facilitate a comparison with WRF-Urban simulation, the collected data was initially interpolated to create a spatially continuous map\cite{Shandas2019} and then aggregated to a 1 km² grid. Figure 7 spatially compares the observed and modeled T2M between 3-4 PM. The corresponding validation statistics in Table 6 show that the RMSE for UT-GLOBUS was 0.53 K, while it was 1.21 K for the LCZ-based approach. Thus, using UT-GLOBUS UCPs resulted in a better representation of T2M compared to the LCZ approach.

\textbf{Quantification of the efficiency of heat mitigation strategies using UT-GLOBUS:} Simulations were also carried out to assess the effectiveness of heat mitigation strategies during an ongoing heatwave event in Chicago. The heat mitigation strategies simulated included cool roofs, green roofs, and cool pavement. These simulations were compared to a control case where no heat mitigation intervention was applied. The simulations were conducted using UT-GLOBUS and LCZ-based default UCPs. For the cool roof experiment, the albedo of the roofs was modified to 0.8 for all the urban grids. In the green roof experiment, grass plantation was chosen for 60\% of the buildings in each urban grid, and irrigation was scheduled from 9 to 10 PM. As for the cool pavement experiment, the albedo of the streets was increased to 0.3 in all urban grids. Since city-scale experiments on heat mitigation strategies have not been conducted, there is no observational data for comparison. Thus, the goal here is to demonstrate the effect of UCPs on T2M when simulating heat mitigation strategies.

Figure 8a shows the mean T2M between 2-5 PM on August 24, 2021, for the control and cool roof experiment simulations. The objective here is to compare the spatial effect of cool roofs on T2M using UT-GLOBUS and default LCZ-based UCPs, as the diurnal variation of T2M has been previously studied\cite{Tan2023}. Cool roofs reflect more shortwave radiation into the atmosphere compared to conventional roofs, altering the net radiation in the urban canyon and causing cooling. T2M reduction is more effective for urban blocks with low-rise buildings and high \(\lambda_p\) because the fraction of area available for cool roof implementation is large and the roof is close to the pedestrian level\cite{Zonato2021}. Since UT-GLOBUS includes building height and \(\lambda_p\), as shown in Figure 8b, its simulations are expected to be more realistic than those using the LCZ approach. From Figures 8a and 8b, it can be observed that higher cooling occurs where the plan area fraction is higher, implying a higher area availability for cool roof implementation. The spatial plots for the green roof and cool pavement experiments, along with the accompanying discussion, are provided in section 8 of the supplementary material.

\textbf{AC energy consumption and Photovoltaic (PV) energy generation simulation with UT-GLOBUS:} This experiment aims to study the fraction of PV energy generation that can be effectively utilized for cooling purposes using the UT-GLOBUS dataset focuses on Austin, Texas. It should be noted, however, that PV can affect the T2M, which in turn affects cooling energy demand.

Figure 9 shows the daily mean AC energy consumption and PV energy generation on 7th August 2021 (clear-sky day) using UT-GLOBUS and LCZ-based UCPs. The results depicted in Figure 9 demonstrate that the simulations conducted with UT-GLOBUS capture the spatial variability in AC energy consumption, highlighting higher energy consumption in downtown Austin. This observation aligns with the presence of taller buildings and a denser building configuration. On the other hand, the LCZ approach produces a more homogeneous distribution of AC energy consumption, as most of Austin is classified as $LCZ_6$ (open low-rise). Similar patterns can be observed in PV energy generation. By examining the fraction of AC energy consumption supplied by PV, UT-GLOBUS demonstrates a realistic estimate compared to the LCZ approach. For instance, in the downtown area, where taller buildings consume a higher amount of energy for cooling, UT-GLOBUS predicts a smaller fraction of AC energy consumption provided by PV. This outcome aligns with the fact that the energy demand for cooling in downtown buildings far exceeds what PV systems can generate based on the available plan area fraction.

\subsubsection*{Street-scale human thermal comfort simulation with UT-GLOBUS}
The street-scale simulation takes into account the detailed urban 3-D geometry, unlike the simplified bulk urban parameterization using UCPs discussed earlier in the city-scale modeling approach. These simulations are important as the bulk parametrizations do not capture the features that are necessary to study local urban hazards such as human urban thermal comfort. 

We conducted thermal comfort simulations using the SOLWEIG model with UT-GLOBUS buildings and used simulations based on LiDAR-derived buildings as a control experiment. Urban trees were not considered in this experiment. The LiDAR-derived buildings were converted to Level of Detail-1 for simulation. We focused on downtown Baltimore and the surrounding residential area, simulating a clear-sky day (August 29, 2019). For the simulation, we used near-surface meteorological data from the ERA-5 reanalysis. The meteorological variables used in the SOLWEIG model simulation were T2M, relative humidity, wind speed, and downwelling shortwave radiation. 

Figure 10 illustrates the simulation of daytime and nighttime mean $T_{MRT}$ values with buildings masked out, using both UT-GLOBUS and LiDAR-derived buildings. During the daytime, downtown Baltimore exhibits lower $T_{MRT}$ values compared to nearby residential areas, while at nighttime, downtown experiences higher $T_{MRT}$ values. This pattern is due to the shadows cast by taller buildings during the day and the reduced sky view factor in downtown areas, which traps longwave radiation from building walls and streets, releasing it at night. The RMSE values for $T_{MRT}$ are 2.85°C during the daytime and 0.9°C at night when compared to LiDAR-based simulations. The difference between $T_{MRT}$ simulations using UT-GLOBUS and LiDAR-based building heights shows that UT-GLOBUS could capture $T_{MRT}$ values within ±5°C during the day and ±2.5°C at night for most of the simulation domain. Thus, based on the RMSE, UT-GLOBUS demonstrated satisfactory performance for thermal comfort simulations.

\subsubsection*{Limitations and uncertainty}
The primary purpose of UT-GLOBUS is to provide a framework to enable modelers to incorporate realistic urban morphology into their simulations to assess urban- and bio-meteorology. The uncertainties in UT-GLOBUS data arise due to multiple factors such as the spatial resolution and uncertainty of ALOS and WSF 3-D datasets, uncertainty and urban coverage of space-borne altimeters, missing building footprints, and the assumption regarding linear population correction factor. It is important to be aware of these limitations and uncertainties when using UT-GLOBUS data for modeling and analysis. Despite these limitations, UT-GLOBUS provides adequately accurate data for urban microclimate modeling as demonstrated by the applications shown.

\section*{Code availability}
GlobalMapper is available at this GitHub repository: \url{https://github.com/Arking1995/GlobalMapper}. The ALOS, GEDI, Google building footprints, and ESA land cover datasets were obtained through Google Earth Engine, while ICESat-2 data was acquired using the icepyx Python package. Microsoft and OSM building data were downloaded using APIs. ICESat-2 data was processed using PhoREAL (\url{https://github.com/icesat-2UT/PhoREAL}) and LAStools (\url{https://lastools.github.io/}). All the datasets were processed using QGIS for Python version 3.26. The scripts for data downloading and data processing can be found at the UT-GLOBUS GitHub repository (\url{https://github.com/texuslabut/UT-GLOBUS}). The urban canopy parameters were converted to WRF binary file format using a Fortran code provided in the UT-GLOBUS GitHub repository.

\section*{Acknowledgements} 
Study benefited from the following research grants: US National Science Foundation (NSF) awards 2324744, 2413827, 1835739, National Aeronautics and Space Administration (NASA) under Interdisciplinary Science (IDS)  80NSSC20K1262, 80NSSC20K1268, US Department of Energy (DoE) ASCR  DE-SC0022211, and  Urban Integrated Field Lab Community Research on Climate and Urban Science (CROCUS) DE-SC0023226, and NOAA Extreme Heat NA21OAR4310146. D.N and Z-L.Y acknowledge Jackson School of Geosciences Farish and Jackson Chair endowments, as well as the Oden Institute of Computational Science and Engineering Visiting Fellowship at UT Austin.

\section*{Author contributions statement}
H.K. led the project. H.K., L.M., and D.N conceived the idea. The data was generated by H.K., M.S., N.M., L.H, and D.A. GlobalMapper is the work of L.H., and D.A. UT-GLOBUS was interfaced with WRF urban by H.K., A.M., C.H.and F.C. H.K. conducted the experiment(s) and verification(s), H.K., A.M., L.M., Z-L.Y., D.N. analyzed the results. H.K. wrote the manuscript draft with support from all authors. Funding for the different projects was obtained and managed by D.N. All authors reviewed, edited, and approved the manuscript. 

\section*{Competing interests} 
The authors declare no competing interests.

\section*{Figures \& Tables}

\begin{table}[h!]
    \centering
    \caption{UT-GLOBUS derived urban canopy parameters (UCPs) for urban canopy models that are implemented in WRF-Urban.}
    \begin{tabular}{cc}
        \toprule
        \textbf{UT-GLOBUS derived UCPs} & \textbf{Used by UCM} \\ 
        \midrule
        Plan area fraction $(\lambda_p)$ & \\
        Area averaged building height $(h_a)$ & Both single-layer and multi-layer UCMs \\
        Building surface to plan area ratio $(\lambda_b)$ & \\
        \midrule
        Histogram of building heights & Multi-layer UCM \\
        \midrule
        Mean building heights & \\
        Standard deviation of building heights & Single-layer UCM \\
        Frontal area index & \\
        \bottomrule
    \end{tabular}
\end{table}

\begin{table}[h!]
\centering
\caption{Comparison of UT-GLOBUS validation statistics with ground truth and existing datasets. Validation statistics used are root mean squared error (RMSE), mean bias error (MBE), coefficient of determination (R²), and spatial correlation coefficient (SC).}
\begin{tabular}{ccccccc}
\toprule
 \textbf{City} &  \textbf{Dataset} &  \textbf{Ground truth} &  \textbf{RMSE (m)} &  \textbf{MBE (m)} &  \textbf{\(R^2\)} &  \textbf{SC} \\ 
 \midrule
Austin, USA & UT-GLOBUS & & 3.85 & -0.22 & 0.84 & 0.92 \\
 & Li et al., (2020) & & 4.49 & 1 & 0.41 & 0.64 \\

Chicago, USA & UT-GLOBUS & & 13.38 & -4.37 & 0.94 & 0.97 \\
 & Li et al., (2020) & & 31.47 & 18.66 & 0.59 & 0.76 \\

Houston, USA & UT-GLOBUS & USGS 3DEP & 9.69 & -1.13 & 0.71 & 0.84 \\
 & Li et al., (2020) & LiDAR & 14.08 & 6.11 & 0.45 & 0.66 \\

Pittsburgh, USA & UT-GLOBUS & & 7.67 & -2.28 & 0.85 & 0.92 \\
 & Li et al., (2020) & & 9.07 & 2.59 & 0.55 & 0.74 \\

Atlanta, USA & UT-GLOBUS & & 8.92 & -2.2 & 0.81 & 0.88 \\
 & Li et al., (2020) & & 9.55 & 1.01 & 0.41 & 0.64 \\

San Antonio, USA & UT-GLOBUS & & 3.23 & -0.09 & 0.74 & 0.86 \\
 & Li et al., (2020) & & 5.8 & 2.54 & 0.44 & 0.66 \\
 \midrule
Hamburg, Germany & UT-GLOBUS & ESA & 4.34 & -2.54 & 0.75 & 0.86 \\
 & Frantz et al., (2021) & & 3.26 & -2.6 & 0.6 & 0.77 \\
 \midrule
Sydney, Australia & UT-GLOBUS & Lipson et al., (2022) & 4.5 & -1.98 & 0.49 & 0.7 \\
\bottomrule
\end{tabular}
\end{table}

\begin{table}[h!]
\centering
\caption{Root mean squared error (RMSE) and spatial correlation coefficient (SC) between UT-GLOBUS and ground truth urban canopy parameters (UCPs) that are shown in Figure 5 for Madrid, Spain.}
\begin{tabular}{ccccc}
\hline
 \textbf{Urban canopy parameters (UCPs)} & \textbf{RMSE} & \textbf{MBE} & \textbf{R²} & \textbf{SC} \\ 
\hline
Plan area fraction ($\lambda_p$) & 0.06 & -0.1 & 0.63 & 0.79 \\
Averaged building heights ($h_a$) & 5.53 (m) & 4.3 (m) & 0.72 & 0.85 \\
Building surface to plan area ratio ($\lambda_b$) & 0.21 & 0.1 & 0.64 & 0.8 \\
\bottomrule
\end{tabular}
\end{table}

\begin{table}[ht]
    \centering
    \caption{City-scale WRF-Urban experiments. Experiments were conducted using UT-GLOBUS  and default local climate zone (LCZ) based urban canopy parameters (UCPs). Simulations using LCZ-based UCPs were used as control cases. LST: land surface temperature, T2M: air temperature, AC: air conditioner, 
    PV: photovoltaic.}
    \begin{tabularx}{\textwidth}{cccccc}
        \toprule
         \textbf{City} &  \textbf{Dates} &  \textbf{Case} &  \textbf{Reason} \\
        \midrule
        Chicago & 21-26 August, 2021 & Heatwave &  Spatial comparison of LST and heat mitigation strategies \\
        Houston & 6-8 August, 2020 & Clear-sky and calm winds &  Spatial comparison of LST and T2M \\
        Austin & 6-8 August, 2020 & Clear-sky and calm winds &  AC energy consumption and PV energy consumption \\
        \bottomrule
    \end{tabularx}
\end{table}

\begin{table}[h!]
\centering
\caption{Statistical comparison of UT-GLOBUS simulated and ECOSTRESS land surface temperature (LST) for Chicago (12 PM) and Houston (12 AM).}
\begin{tabular}{cccccc}
\hline
\textbf{City} & \textbf{Dataset} & \textbf{RMSE (K)} & \textbf{MBE (K)} & \textbf{R²} & \textbf{Spatial correlation (SC)} \\ 
\hline
Chicago, USA & UT-GLOBUS & \text{2.27} & \text{1.23} & 0.16 & 0.4 \\
 & LCZ & 3.03 & 2.03 & 0.01 & 0.05 \\
\hline
Houston, USA & UT-GLOBUS & 1.22 & \text{0.51} & 0.45 & 0.67 \\
 & LCZ & \text{1.1} & -0.66 & \text{0.47} & \text{0.69} \\
\hline
\end{tabular}
\end{table}

\begin{table}[h!]
\centering
\caption{Statistical comparison of UT-GLOBUS simulated and measured air temperature (T2M) for Houston between 3-4 PM on August 7\textsuperscript{th}, 2020.}
\begin{tabular}{cccccc}
\hline
\textbf{City} & \textbf{Dataset} & \textbf{RMSE (K)} & \textbf{MBE (K)} & \textbf{R²} & \textbf{Spatial correlation} \\ 
\hline
Houston, USA & UT-GLOBUS & \text{0.53} & 0 & \text{0.33} & \text{0.58} \\
 & LCZ & 1.21 & -1.05 & 0.26 & 0.27 \\
\hline
\end{tabular}
\end{table}

\begin{figure}[ht]
\centering
\includegraphics[width=\linewidth]{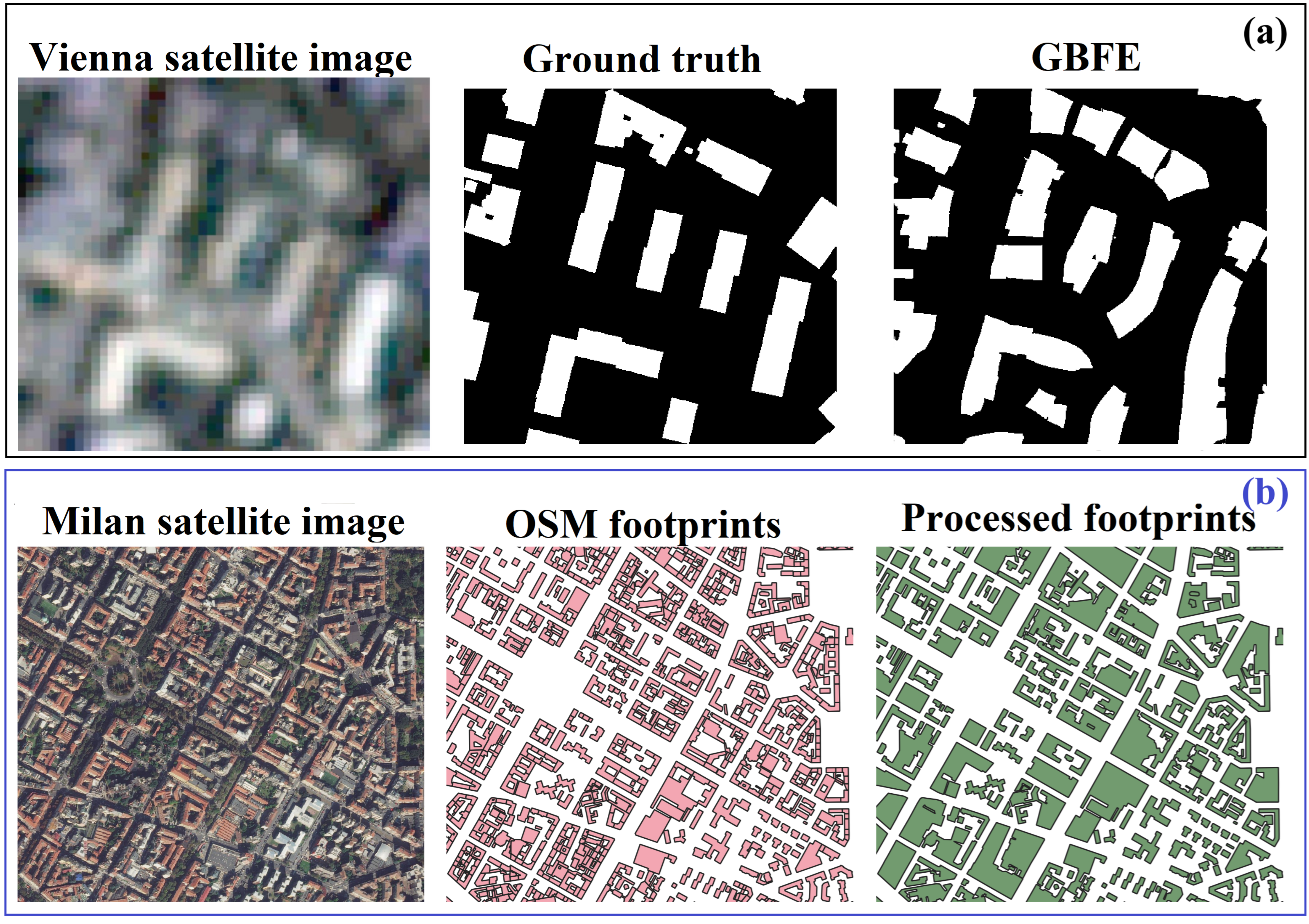}
\caption{(a) Comparison of building footprints derived using GBFE from 3-meter resolution satellite images with ground truth for Vienna, Austria, and (b) Illustration of OSM building footprints processed for Milan, Italy. GBFE: generative building feature estimation and OSM: OpenStreetMaps.}
\label{fig:stream}
\end{figure}

\begin{figure}[ht]
\centering
\includegraphics[width=\linewidth]{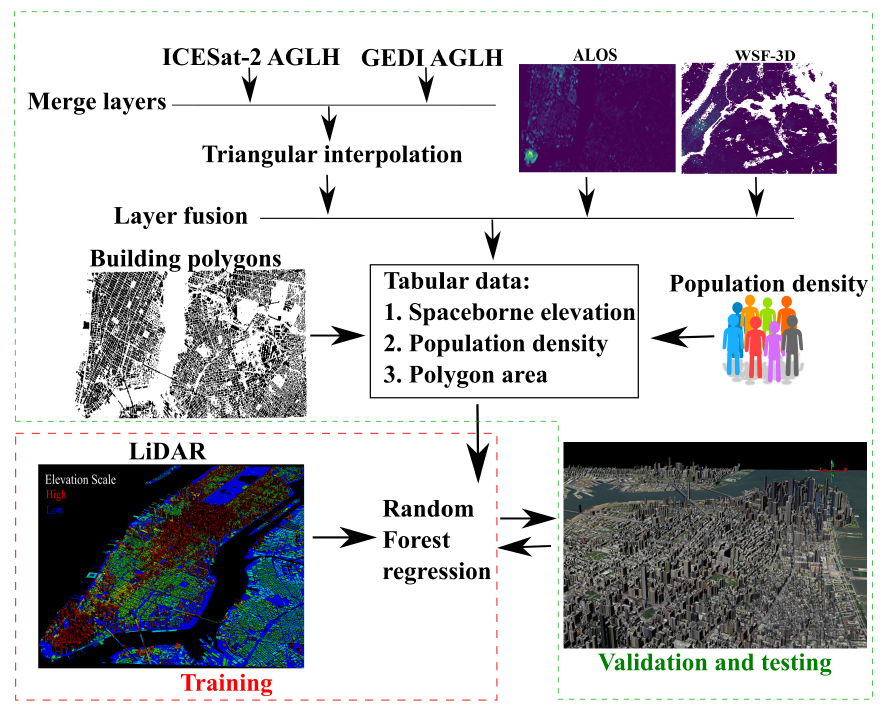}
\caption{UT-GLOBUS methodology for predicting and validating building heights. AGLH: Above ground-level height, ICESat-2: Ice, Cloud, and land Elevation Satellite-2, GEDI: Global Ecosystem Dynamics Investigation, ALOS: Advanced Land Observation Satellite, LiDAR: Light Detection and Ranging.}
\label{fig:stream}
\end{figure}

\begin{figure}[ht]
\centering
\includegraphics[width=\linewidth]{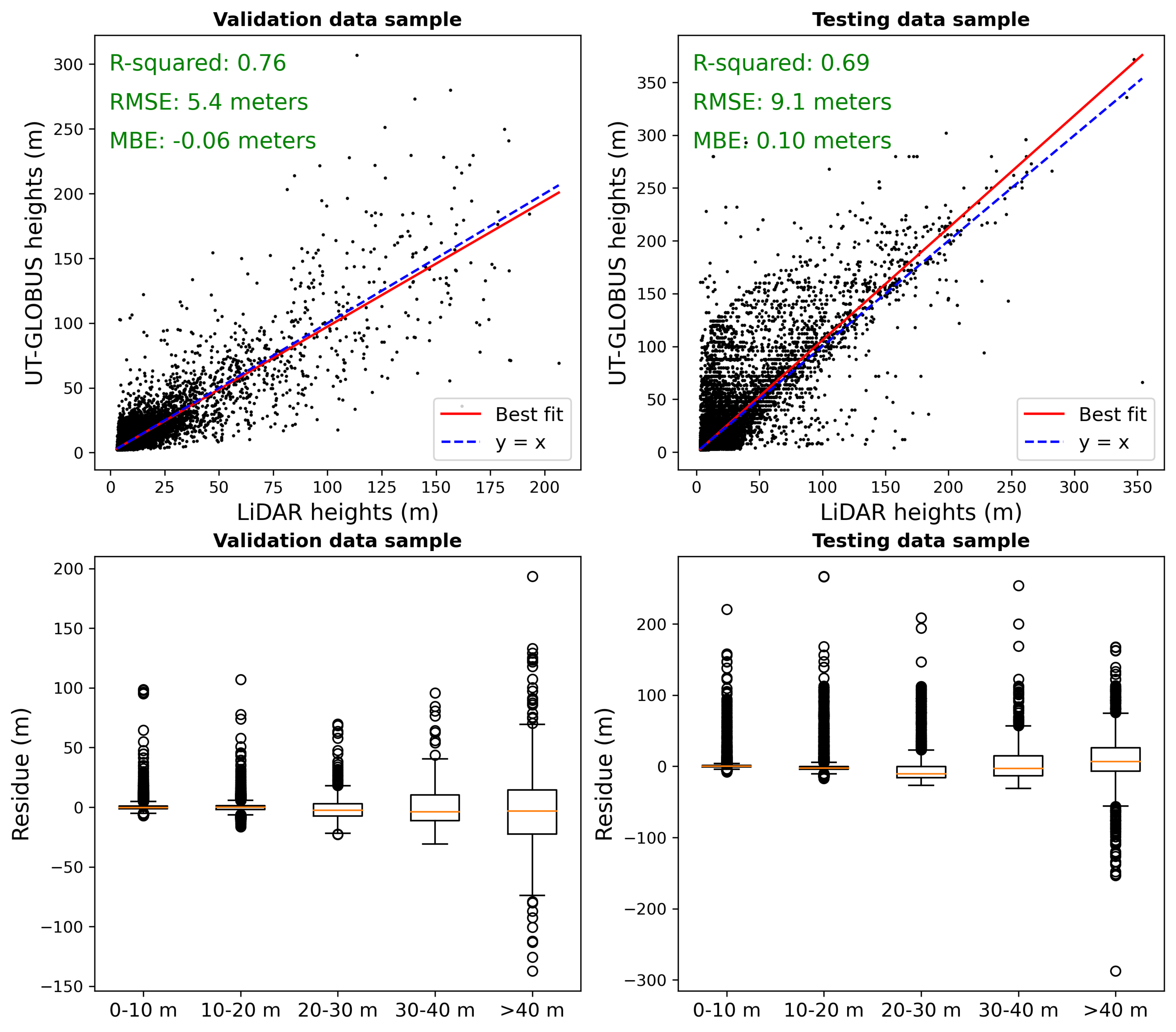}
\caption{Scatter plots (top row) showing the agreement of the random forest model prediction of individual building heights with ground truth for the validation and testing datasets. Box plots (bottom row) present the residues from prediction for \(10\)-meter height bins.} 
\label{fig:stream}
\end{figure}

\begin{figure}[ht]
\centering
\includegraphics[width=\linewidth]{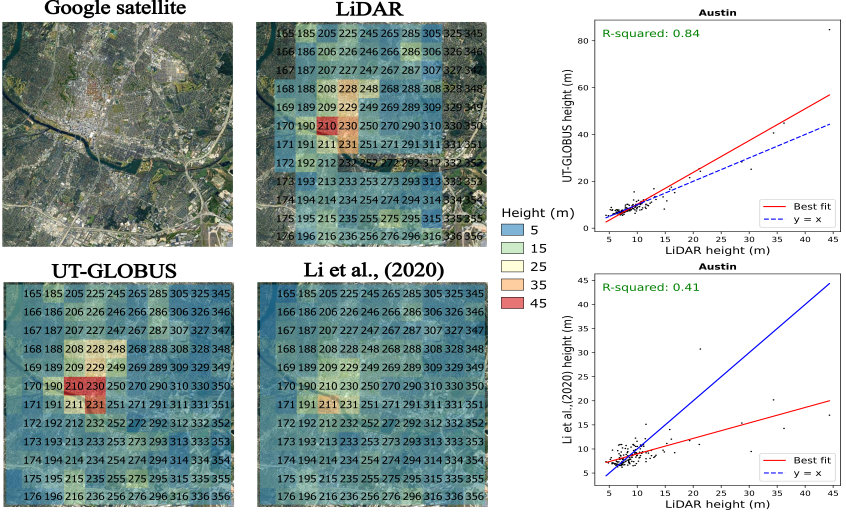}
\caption{Spatial comparison of UT-GLOBUS building heights at 1 km$^2$ resolution against LiDAR (ground truth) and Li et al., (2020) for Austin, Texas. The figure also shows the scatter plots comparing the LiDAR-derived building heights with UT-GLOBUS and Li et al., (2020) datasets.} 
\label{fig:stream}
\end{figure}

\begin{figure}[ht]
\centering
\includegraphics[width=\linewidth]{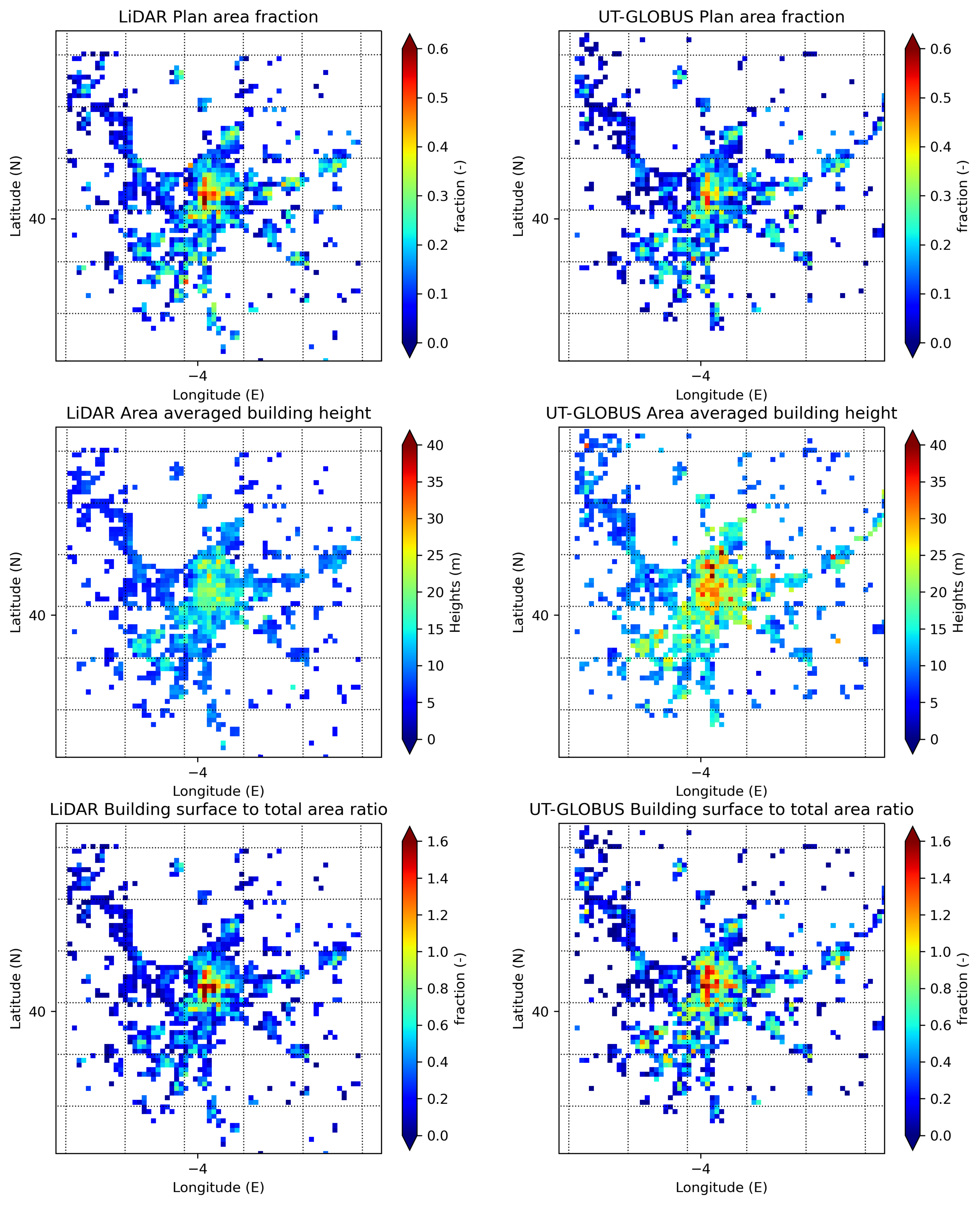}
\caption{Spatial comparison of UT-GLOBUS plan area fraction (\(\lambda_p\)), averaged building heights (\(h_a\)) and building surface to plan area ratio (\(\lambda_b\)) at 1 km$^2$ spatial resolution against Spanish building inventory dataset (ground truth).} 
\label{fig:stream}
\end{figure}

\begin{figure}[ht]
\centering
\includegraphics[width=\linewidth]{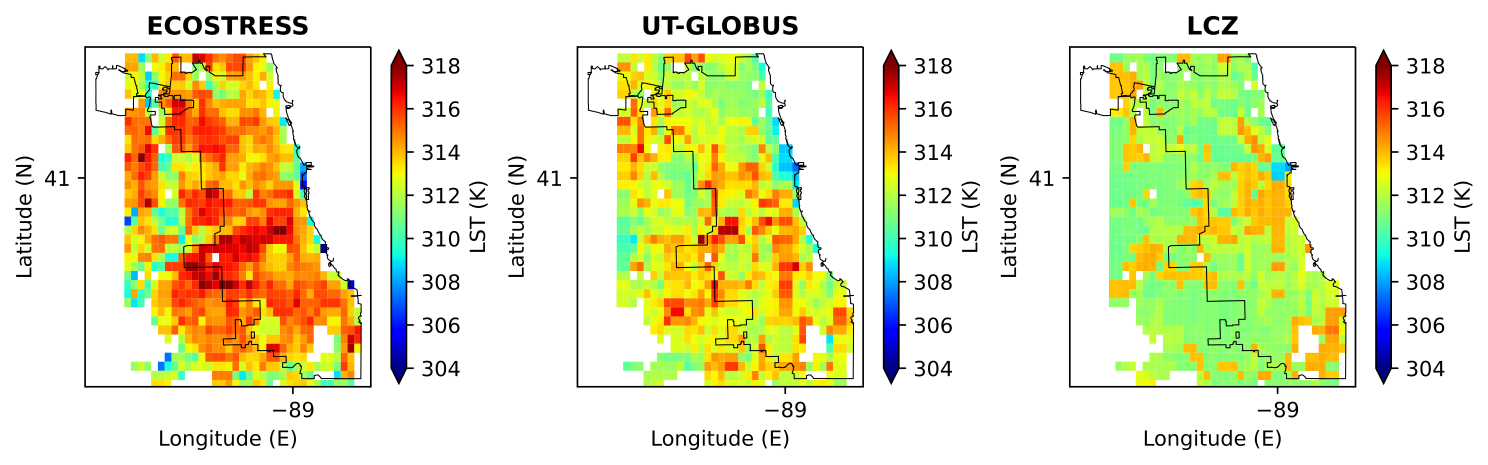}
\caption{WRF-Urban simulated land surface temperature (LST) comparison against ECOSTRESS using UT-GLOBUS and local climate zone-based urban canopy parameters for Chicago at 12 PM.} 
\label{fig:stream}
\end{figure}

\begin{figure}[ht]
\centering
\includegraphics[width=\linewidth]{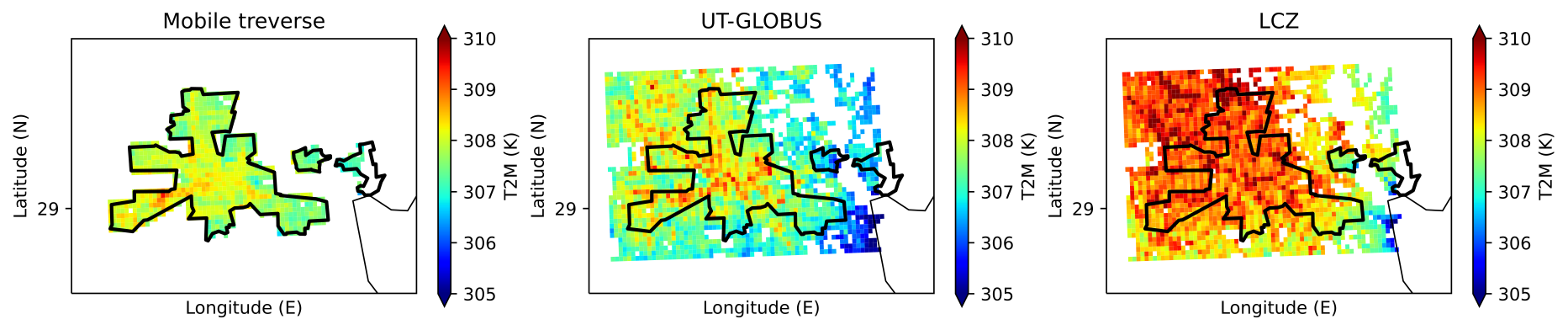}
\caption{Spatial comparison of measured and modeled 2-meter air temperature (T2M) for Houston between 3-4 PM using UT-GLOBUS and local climate zone-based urban canopy parameters} 
\label{fig:stream}
\end{figure}

\begin{figure}[ht]
\centering
\includegraphics[width=\linewidth]{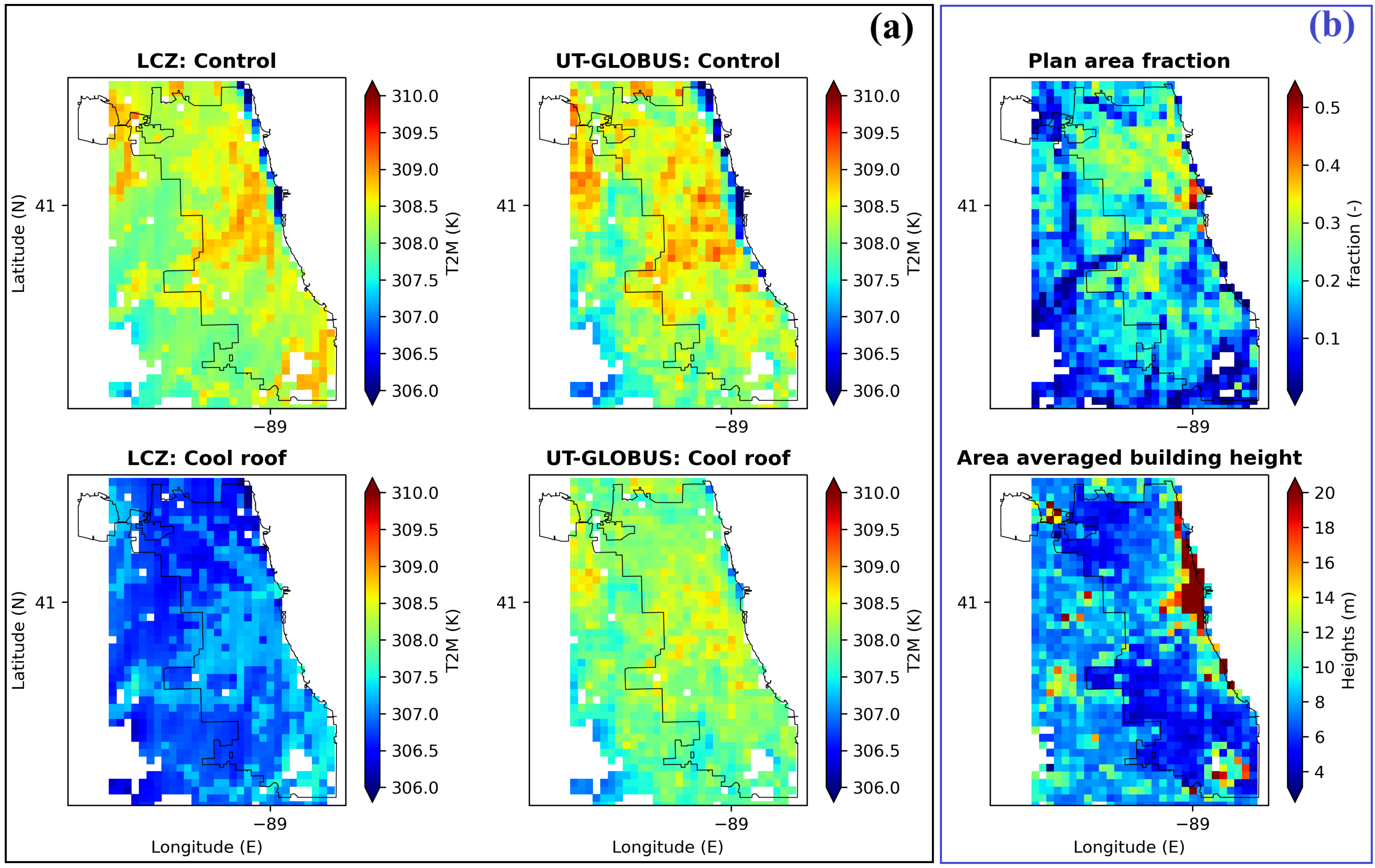}
\caption{(a) Mean air temperature between 2-5 PM on August 24th, 2021, for control and cool roof heat mitigation strategy simulations using UT-GLOBUS and local climate zone-based urban canopy parameters (UCPs) for Chicago (b) UCPs for Chicago: Plan area fraction and area-averaged building heights.} 
\label{fig:stream}
\end{figure}

\begin{figure}[ht]
\centering
\includegraphics[width=\linewidth]{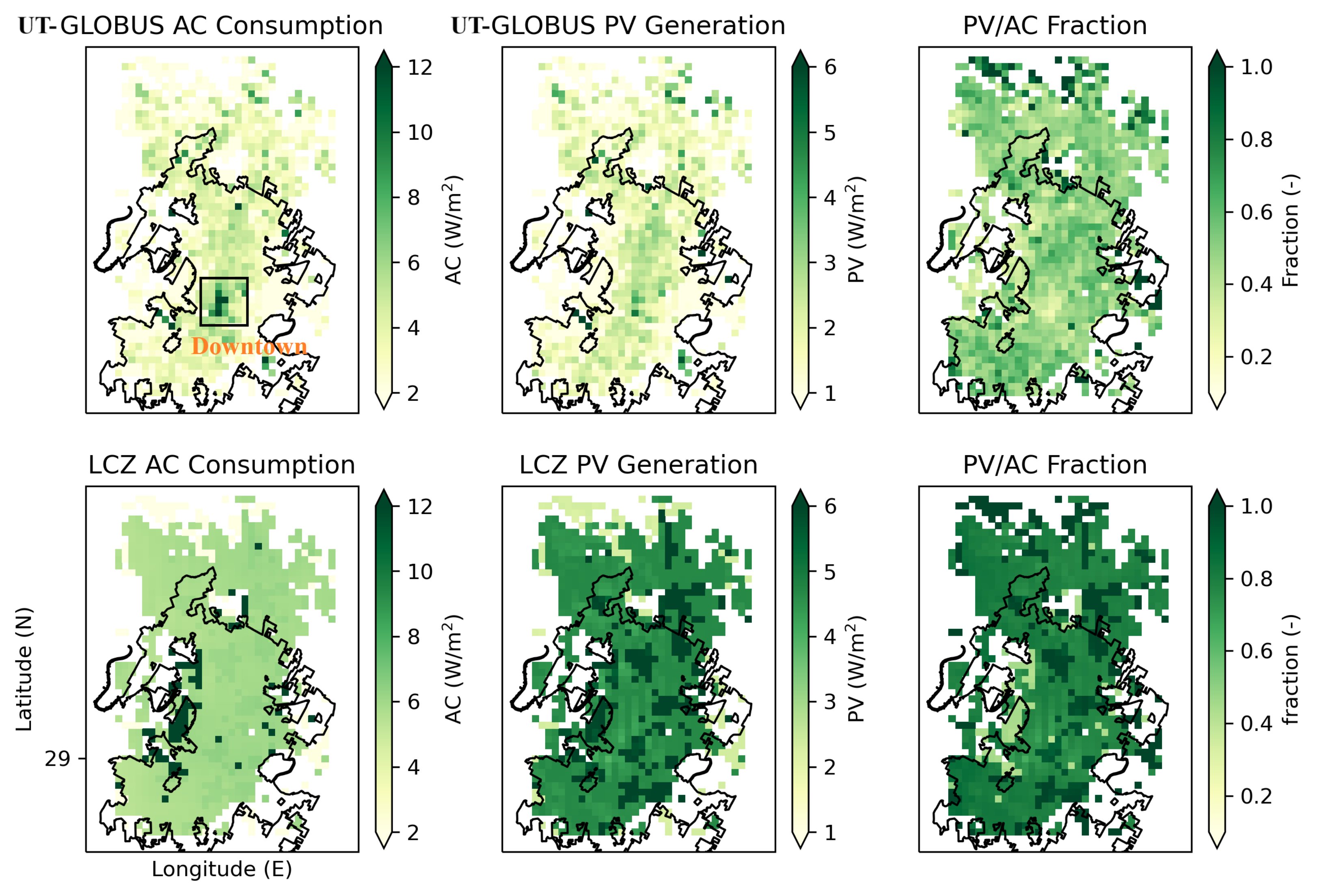}
\caption{Daily mean AC energy consumption and PV energy generation using UT-GLOBUS and local climate zone based urban canopy parameters for Austin, Texas. The downtown area is highlighted by a box in the first panel.} 
\label{fig:stream}
\end{figure}

\begin{figure}[ht]
\centering
\includegraphics[width=\linewidth]{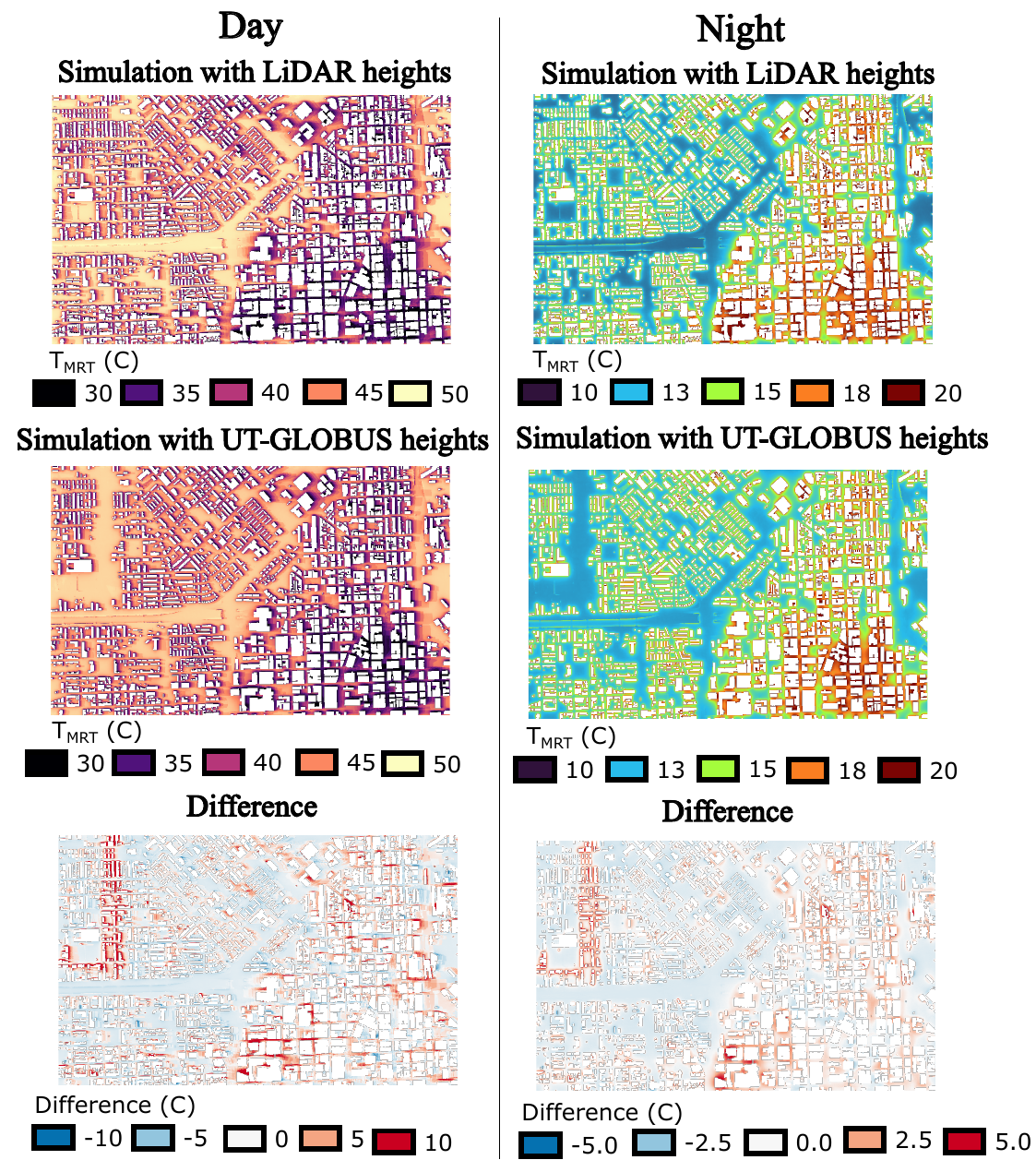}
\caption{Daytime and Nighttime mean $T_{MRT}$ for downtown Baltimore and surrounding residential area using LiDAR and UT-GLOBUS buildings.} 
\label{fig:stream}
\end{figure}


\begin{thebibliography}{10}
\urlstyle{rm}
\expandafter\ifx\csname url\endcsname\relax
  \def\url#1{\texttt{#1}}\fi
\expandafter\ifx\csname urlprefix\endcsname\relax\def\urlprefix{URL }\fi
\expandafter\ifx\csname doiprefix\endcsname\relax\def\doiprefix{DOI: }\fi
\providecommand{\bibinfo}[2]{#2}
\providecommand{\eprint}[2][]{\url{#2}}

\bibitem{kamath2023human}
\bibinfo{author}{Kamath, H.~G.} \emph{et~al.}
\newblock \bibinfo{journal}{\bibinfo{title}{Human heat health index (h3i) for holistic assessment of heat hazard and mitigation strategies beyond urban heat islands}}.
\newblock {\emph{\JournalTitle{Urban Climate}}} \textbf{\bibinfo{volume}{52}}, \bibinfo{pages}{101675} (\bibinfo{year}{2023}).

\bibitem{lewis2024fire}
\bibinfo{author}{Lewis, R.~H.} \emph{et~al.}
\newblock \bibinfo{journal}{\bibinfo{title}{Fire and smoke digital twin--a computational framework for modeling fire incident outcomes}}.
\newblock {\emph{\JournalTitle{Computers, Environment and Urban Systems}}} \textbf{\bibinfo{volume}{110}}, \bibinfo{pages}{102093} (\bibinfo{year}{2024}).

\bibitem{Wang2021}
\bibinfo{author}{Wang, Y.}, \bibinfo{author}{Zorzi, S.} \& \bibinfo{author}{Bittner, K.}
\newblock \bibinfo{title}{Machine-learned 3d building vectorization from satellite imagery}.
\newblock In \emph{\bibinfo{booktitle}{IEEE Computer Society Conference on Computer Vision and Pattern Recognition Workshops}}, \bibinfo{pages}{1072–1081}, \url{https://doi.org/10.1109/CVPRW53098.2021.00118} (\bibinfo{publisher}{IEEE}, \bibinfo{year}{2021}).

\bibitem{Chen2011}
\bibinfo{author}{Chen, F.} \emph{et~al.}
\newblock \bibinfo{journal}{\bibinfo{title}{The integrated wrf/urban modelling system: Development, evaluation, and applications to urban environmental problems}}.
\newblock {\emph{\JournalTitle{International Journal of Climatology}}} \textbf{\bibinfo{volume}{31}}, \bibinfo{pages}{273–288}, \url{https://doi.org/10.1002/joc.2158} (\bibinfo{year}{2011}).

\bibitem{sun2019python}
\bibinfo{author}{Sun, T.} \& \bibinfo{author}{Grimmond, S.}
\newblock \bibinfo{journal}{\bibinfo{title}{A python-enhanced urban land surface model supy (suews in python, v2019. 2): development, deployment and demonstration}}.
\newblock {\emph{\JournalTitle{Geoscientific Model Development}}} \textbf{\bibinfo{volume}{12}}, \bibinfo{pages}{2781--2795} (\bibinfo{year}{2019}).

\bibitem{Lindberg2008}
\bibinfo{author}{Lindberg, F.}, \bibinfo{author}{Holmer, B.} \& \bibinfo{author}{Thorsson, S.}
\newblock \bibinfo{journal}{\bibinfo{title}{Solweig 1.0 - modelling spatial variations of 3d radiant fluxes and mean radiant temperature in complex urban settings}}.
\newblock {\emph{\JournalTitle{Int J Biometeorol}}} \textbf{\bibinfo{volume}{52}}, \bibinfo{pages}{697–713}, \url{https://doi.org/10.1007/s00484-008-0162-7} (\bibinfo{year}{2008}).

\bibitem{Stewart2012}
\bibinfo{author}{Stewart, I.} \& \bibinfo{author}{Oke, T.}
\newblock \bibinfo{journal}{\bibinfo{title}{Local climate zones for urban temperature studies}}.
\newblock {\emph{\JournalTitle{Bull Am Meteorol Soc}}} \textbf{\bibinfo{volume}{93}}, \bibinfo{pages}{1879–1900}, \url{https://doi.org/10.1175/BAMS-D-11-0019.1} (\bibinfo{year}{2012}).

\bibitem{Patel2023}
\bibinfo{author}{Patel, P.}, \bibinfo{author}{Kalyanam, R.}, \bibinfo{author}{He, L.}, \bibinfo{author}{Aliaga, D.} \& \bibinfo{author}{Niyogi, D.}
\newblock \bibinfo{journal}{\bibinfo{title}{Deep learning-based urban morphology for city-scale environmental modeling}}.
\newblock {\emph{\JournalTitle{PNAS Nexus 2}}} \url{https://doi.org/10.1093/pnasnexus/pgad027} (\bibinfo{year}{2023}).

\bibitem{Ching2009}
\bibinfo{author}{Ching, J.} \emph{et~al.}
\newblock \bibinfo{journal}{\bibinfo{title}{National urban database and access portal tool}}.
\newblock {\emph{\JournalTitle{Bull Am Meteorol Soc}}} \textbf{\bibinfo{volume}{90}}, \bibinfo{pages}{1157–1168}, \url{https://doi.org/10.1175/2009BAMS2675.1} (\bibinfo{year}{2009}).

\bibitem{Ching2018}
\bibinfo{author}{Ching, J.} \emph{et~al.}
\newblock \bibinfo{journal}{\bibinfo{title}{Wudapt: An urban weather, climate, and environmental modeling infrastructure for the anthropocene}}.
\newblock {\emph{\JournalTitle{Bull Am Meteorol Soc}}} \textbf{\bibinfo{volume}{99}}, \bibinfo{pages}{1907–1924}, \url{https://doi.org/10.1175/BAMS-D-16-0236.1} (\bibinfo{year}{2018}).

\bibitem{Salamanca2011}
\bibinfo{author}{Salamanca, F.}, \bibinfo{author}{Martilli, A.}, \bibinfo{author}{Tewari, M.} \& \bibinfo{author}{Chen, F.}
\newblock \bibinfo{journal}{\bibinfo{title}{A study of the urban boundary layer using different urban parameterizations and high-resolution urban canopy parameters with wrf}}.
\newblock {\emph{\JournalTitle{J Appl Meteorol Climatol}}} \textbf{\bibinfo{volume}{50}}, \bibinfo{pages}{1107–1128}, \url{https://doi.org/10.1175/2010JAMC2538.1} (\bibinfo{year}{2011}).

\bibitem{Kusaka2001}
\bibinfo{author}{Kusaka, H.}, \bibinfo{author}{Kondo, H.}, \bibinfo{author}{Kikegawa, Y.} \& \bibinfo{author}{Kimura, F.}
\newblock \bibinfo{journal}{\bibinfo{title}{A simple single-layer urban canopy model for atmospheric models: Comparison with multi-layer and slab models}}.
\newblock {\emph{\JournalTitle{Boundary Layer Meteorol}}} \textbf{\bibinfo{volume}{101}}, \bibinfo{pages}{329–358}, \url{https://doi.org/10.1023/A:1019207923078} (\bibinfo{year}{2001}).

\bibitem{Martilli2002}
\bibinfo{author}{Martilli, A.}, \bibinfo{author}{Clappier, A.} \& \bibinfo{author}{Rotach, M.}
\newblock \bibinfo{journal}{\bibinfo{title}{An urban surface exchange parameterisation for mesoscale models}}.
\newblock {\emph{\JournalTitle{Boundary Layer Meteorol}}} \textbf{\bibinfo{volume}{104}}, \bibinfo{pages}{261–304}, \url{https://doi.org/10.1023/A:1016099921195} (\bibinfo{year}{2002}).

\bibitem{Salamanca2010}
\bibinfo{author}{Salamanca, F.} \& \bibinfo{author}{Martilli, A.}
\newblock \bibinfo{journal}{\bibinfo{title}{A new building energy model coupled with an urban canopy parameterization for urban climate simulations-part ii. validation with one dimension off-line simulations}}.
\newblock {\emph{\JournalTitle{Theor Appl Climatol}}} \textbf{\bibinfo{volume}{99}}, \bibinfo{pages}{345–356}, \url{https://doi.org/10.1007/s00704-009-0143-8} (\bibinfo{year}{2010}).

\bibitem{Papaccogli2021}
\bibinfo{author}{Papaccogli, G.}, \bibinfo{author}{Giovannini, L.}, \bibinfo{author}{Zardi, D.} \& \bibinfo{author}{Martilli, A.}
\newblock \bibinfo{journal}{\bibinfo{title}{Assessing the ability of wrf-bep + bem in reproducing the wintertime building energy consumption of an italian alpine city}}.
\newblock {\emph{\JournalTitle{Journal of Geophysical Research: Atmospheres}}} \textbf{\bibinfo{volume}{126}}, \url{https://doi.org/10.1029/2020JD033652} (\bibinfo{year}{2021}).

\bibitem{Zonato2021}
\bibinfo{author}{Zonato, A.} \emph{et~al.}
\newblock \bibinfo{journal}{\bibinfo{title}{Exploring the effects of rooftop mitigation strategies on urban temperatures and energy consumption}}.
\newblock {\emph{\JournalTitle{Journal of Geophysical Research: Atmospheres}}} \textbf{\bibinfo{volume}{126}}, \bibinfo{pages}{1–24}, \url{https://doi.org/10.1029/2021jd035002} (\bibinfo{year}{2021}).

\bibitem{Tan2023}
\bibinfo{author}{Tan, H.}, \bibinfo{author}{Kotamarthi, R.}, \bibinfo{author}{Wang, J.}, \bibinfo{author}{Qian, Y.} \& \bibinfo{author}{Chakraborty, T.}
\newblock \bibinfo{journal}{\bibinfo{title}{Impact of different roofing mitigation strategies on near-surface temperature and energy consumption over the chicago metropolitan area during a heatwave event}}.
\newblock {\emph{\JournalTitle{Science of the Total Environment}}} \textbf{\bibinfo{volume}{860}}, \url{https://doi.org/10.1016/j.scitotenv.2022.160508} (\bibinfo{year}{2023}).

\bibitem{Bernard2022}
\bibinfo{author}{Bernard, J.}, \bibinfo{author}{Bocher, E.}, \bibinfo{author}{Le, E.}, \bibinfo{author}{Wiederhold, S.} \& \bibinfo{author}{Leconte, F.}
\newblock \bibinfo{journal}{\bibinfo{title}{Estimation of missing building height in openstreetmap data: a french case study using geoclimate 0.0.1}}.
\newblock {\emph{\JournalTitle{Geoscientific Model Development Discussions}}} \url{https://doi.org/10.5194/gmd-2021-428} (\bibinfo{year}{2022}).

\bibitem{Reichstein2019}
\bibinfo{author}{Reichstein, M.} \emph{et~al.}
\newblock \bibinfo{journal}{\bibinfo{title}{Deep learning and process understanding for data-driven earth system science}}.
\newblock {\emph{\JournalTitle{Nature}}} \textbf{\bibinfo{volume}{566}}, \bibinfo{pages}{195–204}, \url{https://doi.org/10.1038/s41586-019-0912-1} (\bibinfo{year}{2019}).

\bibitem{Breiman2001}
\bibinfo{author}{Breiman, L.}
\newblock \emph{\bibinfo{title}{Random Forests}} (\bibinfo{year}{2001}).

\bibitem{Tadono2015}
\bibinfo{author}{Tadono, T.}, \bibinfo{author}{Takaku, J.}, \bibinfo{author}{Tsutsui, K.}, \bibinfo{author}{Oda, F.} \& \bibinfo{author}{Nagai, H.}
\newblock \bibinfo{title}{Status of alos world 3d (aw3d) global dsm generation}.
\newblock In \emph{\bibinfo{booktitle}{2015 IEEE International Geoscience and Remote Sensing Symposium (IGARSS)}}, \bibinfo{pages}{3822–3825}, \url{https://doi.org/10.1109/IGARSS.2015.7326657} (\bibinfo{publisher}{IEEE}, \bibinfo{year}{2015}).

\bibitem{Neuenschwander2019}
\bibinfo{author}{Neuenschwander, A.} \& \bibinfo{author}{Pitts, K.}
\newblock \bibinfo{journal}{\bibinfo{title}{The atl08 land and vegetation product for the icesat-2 mission}}.
\newblock {\emph{\JournalTitle{Remote Sens Environ}}} \textbf{\bibinfo{volume}{221}}, \bibinfo{pages}{247–259}, \url{https://doi.org/10.1016/j.rse.2018.11.005} (\bibinfo{year}{2019}).

\bibitem{esch2022world}
\bibinfo{author}{Esch, T.} \emph{et~al.}
\newblock \bibinfo{journal}{\bibinfo{title}{World settlement footprint 3d-a first three-dimensional survey of the global building stock}}.
\newblock {\emph{\JournalTitle{Remote sensing of environment}}} \textbf{\bibinfo{volume}{270}}, \bibinfo{pages}{112877} (\bibinfo{year}{2022}).

\bibitem{Frantz2021}
\bibinfo{author}{Frantz, D.} \emph{et~al.}
\newblock \bibinfo{journal}{\bibinfo{title}{National-scale mapping of building height using sentinel-1 and sentinel-2 time series}}.
\newblock {\emph{\JournalTitle{Remote Sens Environ}}} \textbf{\bibinfo{volume}{252}}, \url{https://doi.org/10.1016/j.rse.2020.112128} (\bibinfo{year}{2021}).

\bibitem{dobson2000landscan}
\bibinfo{author}{Dobson, J.~E.}, \bibinfo{author}{Bright, E.~A.}, \bibinfo{author}{Coleman, P.~R.}, \bibinfo{author}{Durfee, R.~C.} \& \bibinfo{author}{Worley, B.~A.}
\newblock \bibinfo{journal}{\bibinfo{title}{Landscan: a global population database for estimating populations at risk}}.
\newblock {\emph{\JournalTitle{Photogrammetric engineering and remote sensing}}} \textbf{\bibinfo{volume}{66}}, \bibinfo{pages}{849--857} (\bibinfo{year}{2000}).

\bibitem{he2023generative}
\bibinfo{author}{He, L.}, \bibinfo{author}{Shan, J.} \& \bibinfo{author}{Aliaga, D.}
\newblock \bibinfo{journal}{\bibinfo{title}{Generative building feature estimation from satellite images}}.
\newblock {\emph{\JournalTitle{IEEE Transactions on Geoscience and Remote Sensing}}} \textbf{\bibinfo{volume}{61}}, \bibinfo{pages}{1--13} (\bibinfo{year}{2023}).

\bibitem{he2023globalmapper}
\bibinfo{author}{He, L.} \& \bibinfo{author}{Aliaga, D.}
\newblock \bibinfo{title}{Globalmapper: Arbitrary-shaped urban layout generation}.
\newblock In \emph{\bibinfo{booktitle}{Proceedings of the IEEE/CVF International Conference on Computer Vision}}, \bibinfo{pages}{454--464} (\bibinfo{year}{2023}).

\bibitem{kamath_2024}
\bibinfo{author}{Kamath, H.} \emph{et~al.}
\newblock \bibinfo{title}{Global building heights for urban studies (ut-globus)},
\newblock {\emph{\JournalTitle{Zenodo}}},
\url{https://doi.org/10.5281/zenodo.11156602} (\bibinfo{year}{2024}).

\bibitem{li2020developing}
\bibinfo{author}{Li, X.}, \bibinfo{author}{Zhou, Y.}, \bibinfo{author}{Gong, P.}, \bibinfo{author}{Seto, K.~C.} \& \bibinfo{author}{Clinton, N.}
\newblock \bibinfo{journal}{\bibinfo{title}{Developing a method to estimate building height from sentinel-1 data}}.
\newblock {\emph{\JournalTitle{Remote Sensing of Environment}}} \textbf{\bibinfo{volume}{240}}, \bibinfo{pages}{111705} (\bibinfo{year}{2020}).

\bibitem{lipson2022transformation}
\bibinfo{author}{Lipson, M.~J.}, \bibinfo{author}{Nazarian, N.}, \bibinfo{author}{Hart, M.~A.}, \bibinfo{author}{Nice, K.~A.} \& \bibinfo{author}{Conroy, B.}
\newblock \bibinfo{journal}{\bibinfo{title}{A transformation in city-descriptive input data for urban climate models}}.
\newblock {\emph{\JournalTitle{Frontiers in Environmental Science}}} \textbf{\bibinfo{volume}{10}}, \bibinfo{pages}{866398} (\bibinfo{year}{2022}).

\bibitem{Hulley2022}
\bibinfo{author}{Hulley, G.} \emph{et~al.}
\newblock \bibinfo{journal}{\bibinfo{title}{Validation and quality assessment of the ecostress level-2 land surface temperature and emissivity product}}.
\newblock {\emph{\JournalTitle{IEEE Transactions on Geoscience and Remote Sensing}}} \textbf{\bibinfo{volume}{60}}, \url{https://doi.org/10.1109/TGRS.2021.3079879} (\bibinfo{year}{2022}).

\bibitem{Voogt1997}
\bibinfo{author}{Voogt, J.} \& \bibinfo{author}{Oke, T.}
\newblock \bibinfo{journal}{\bibinfo{title}{Complete urban surface temperatures}}.
\newblock {\emph{\JournalTitle{Journal of Applied Meteorology}}} \textbf{\bibinfo{volume}{36}}, \bibinfo{pages}{1117–1132}, \url{https://doi.org/10.1175/1520-0450(1997)036<1117:CUST>2.0.CO;2} (\bibinfo{year}{1997}).

\bibitem{anderson2021interoperability}
\bibinfo{author}{Anderson, M.~C.} \emph{et~al.}
\newblock \bibinfo{journal}{\bibinfo{title}{Interoperability of ecostress and landsat for mapping evapotranspiration time series at sub-field scales}}.
\newblock {\emph{\JournalTitle{Remote Sensing of Environment}}} \textbf{\bibinfo{volume}{252}}, \bibinfo{pages}{112189} (\bibinfo{year}{2021}).

\bibitem{oke2017urban}
\bibinfo{author}{Oke, T.~R.}, \bibinfo{author}{Mills, G.}, \bibinfo{author}{Christen, A.} \& \bibinfo{author}{Voogt, J.~A.}
\newblock \emph{\bibinfo{title}{Urban climates}} (\bibinfo{publisher}{Cambridge University Press}, \bibinfo{year}{2017}).

\bibitem{Shandas2019}
\bibinfo{author}{Shandas, V.}, \bibinfo{author}{Voelkel, J.}, \bibinfo{author}{Williams, J.} \& \bibinfo{author}{Hoffman, J.}
\newblock \bibinfo{journal}{\bibinfo{title}{Integrating satellite and ground measurements for predicting locations of extreme urban heat}}.
\newblock {\emph{\JournalTitle{Climate}}} \textbf{\bibinfo{volume}{7}}, \url{https://doi.org/10.3390/cli7010005} (\bibinfo{year}{2019}).

\end{thebibliography}
\end{document}